\def\apj{{ApJ}}

\def\mnras{{MNRAS}}

\def\kms{km s$^{-1}$}
\def\lum{erg s$^{-1}$}

\def\farcs{\hbox{$.\mkern-4mu^{\prime\prime}$}}

\documentclass[cup5b]{caps}
\usepackage{graphicx}
\usepackage{amssymb}
\usepackage{ociwsymp1}
\HeadText{R. P. van der Marel}

\begin{document}

\pagenumbering{arabic}

\author[]{R. P. VAN DER MAREL\\Space Telescope Science Institute}

\chapter{Intermediate-Mass Black Holes in 
         the Universe: A Review of Formation 
         Theories and Observational Constraints}

\begin{abstract}
This paper reviews the subject of intermediate-mass black holes
(IMBHs) with masses between those of ``stellar-mass'' and
``supermassive'' black holes (BHs). The existence of IMBHs is a real
possibility: they might plausibly have formed as remnants of the first
generation of stars (Population~III), as the result of dense star
cluster evolution, or as part of the formation process of
supermassive BHs. Their cosmic mass density could exceed that
of supermassive BHs ($\Omega \approx 10^{-5.7}$) and
observations do not even rule out that they may account for all of the
baryonic dark matter in the Universe ($\Omega \approx 10^{-1.7}$).
Unambiguous detections of individual IMBHs currently do not exist, but
there are observational hints from studies of microlensing events,
``ultra-luminous'' X-ray sources, and centers of nearby galaxies and
globular clusters. Gravitational wave experiments will soon provide
another method to probe their existence. IMBHs have potential
importance for several fields of astrophysics and are likely to grow
as a focus of research attention.
\end{abstract}

\section{Introduction}
\label{s:intro}

BHs were long considered a mathematical curiosity, but it is now clear
that they are an important and indisputable part of the astronomical
landscape (e.g., Begelman \& Rees 1998). In particular, there is
unambiguous evidence for ``stellar-mass'' BHs and ``supermassive'' BHs.

Stellar-mass BHs form in a reasonably well-understood manner through
stellar evolution (Fryer 1999). A BH with a companion star
might accrete matter from it to produce an X-ray binary (XRB). It is
sometimes possible to determine the mass of the accreting object in an
XRB from detailed modeling. Neutron stars cannot be more massive than
2--$3\, M_{\odot}$ and compact objects with larger masses are therefore
assumed to be BHs. Some dozen such objects are know, mostly with
masses in the range 5--$15\, M_{\odot}$ (Charles 2001). The fraction of
stellar-mass BHs that manages to remain in a close binary throughout
its evolution and is also currently accreting is very small, so most
stellar-mass BHs exist singly and go unnoticed
(\S~\ref{ss:bulgelensing}). The Milky Way hosts $\sim 10^{7-9}$
stellar-mass BHs (Brown \& Bethe 1994).

The paradigm that there are also supermassive BHs in the Universe is
based on the existence of active galactic nuclei (AGNs) in the centers
of some galaxies. Their properties can only be plausibly explained by
assuming that a BH of $10^6$--$10^9\, M_{\odot}$ acts as the central
engine (Rees 1984). The proper motions of stars around our Galactic
Center (Sgr A$^{*}$) provide direct evidence for this (Sch\"odel et
al. 2002; Ghez 2003). A variety of techniques now exist to detect and weigh
supermassive BHs using stellar or gaseous kinematics (Kormendy \&
Gebhardt 2001). The BH mass is always of the order $0.1$\% of the
galaxy bulge/spheroid mass. An even better correlation exists with the
velocity dispersion, $M_{\rm BH} \propto \sigma^4$ (Tremaine et
al. 2002). The origin of these correlations, the triggers of AGN
activity, and the exact formation mechanisms of supermassive BHs
remain poorly understood (\S~\ref{ss:relatedsupBH}).
 
Stellar-mass and supermassive BHs can be studied because they
(sometimes) exist in environments favorable for the production of
observable signatures through accretion or gravitational
influence. This need not to be true for all BHs in the Universe, which
may therefore also exist in other mass ranges. BHs in the
intermediate-mass range of, say, $15$--$10^6\, M_{\odot}$ (i.e, between
the familiar classes of BHs), are of particular interest. Such IMBHs
might plausibly have formed in different ways
(\S~\ref{s:formation}). They have been suggested as an important
component of the missing baryonic dark matter in the Universe
(\S~\ref{s:darkmatter}) and recent observational studies have
provided hints of IMBHs in various environments
(\S~\ref{s:obsconstraints}). It can be concluded that IMBHs are
an important topic for additional research (\S~\ref{s:conc}).

\section{Formation Theories}
\label{s:formation}

IMBHs may plausibly have formed as the remnants of Population~III
stars (\S~\ref{ss:popIII}), through dynamical processes in dense
star clusters (\S~\ref{ss:dense}), or as an essential ingredient
or occasional by-product of the formation of supermassive BHs
(\S~\ref{ss:relatedsupBH}). Primordial formation of IMBHs is
unlikely (\S~\ref{ss:primordial}).

\subsection{Formation from Population III Stellar Evolution}
\label{ss:popIII}

The present-day stellar initial mass function (IMF) extends to $\sim
200\, M_{\odot}$ (Larson 2003). Massive stars shed most of their mass
through radiatively driven stellar winds. Above $\sim 100\, M_{\odot}$ a
nuclear pulsational instability sets in that generates additional mass
loss. Evolutionary calculations indicate that massive stars leave
compact remnants with masses below $\sim 15\, M_{\odot}$, consistent
with observations of X-ray binaries (Fryer \& Kalogera 2001). The
minimum initial mass for a star to become a stellar-mass BH (rather
than a neutron star) is $\sim 20$--$25\, M_{\odot}$ (Fryer 1999).

For the first generation of zero-metallicity stars in the Universe
(Population III) the initial conditions and evolutionary path were
quite different.  There is now a growing body of evidence that
suggests a top-heavy IMF in the early Universe (Schneider et
al. 2002), although this issue continues to be debated. In the
absence of metals, primordial molecular clouds cool through
rotational-vibrational lines of H$_2$.  Simulations of the
collapse and fragmentation of such clouds (Abel, Bryan, \& Norman 2000)
suggest that the first generation of stars had typical masses of $\sim
100\, M_{\odot}$, compared to the $\sim 1\, M_{\odot}$ characteristic of
stars at the present epoch. In addition, radiative mass losses are
negligible at zero metallicity, and mass losses due to nuclear
pulsational instability are greatly reduced (Fryer, Woosley, \& Heger
2001).\looseness=-2

The evolution of massive Population III stars depends on the initial
mass (Bond, Arnett, \& Carr 1984; Heger \& Woosley 2001). Stars below
$140\, M_{\odot}$ probably evolve into BHs in similar fashion as do
stars of normal metallicity (Fryer 1999), although the remnant BHs will
be more massive than today's stellar-mass BHs due to the more limited
mass loss. Stars that are initially more massive than $\sim 140\,
M_{\odot}$ encounter the electron-positron pair-instability during
oxygen burning. In the range $\sim 140$--$260\, M_{\odot}$ this yields
an explosion that leaves no remnant and is considerably more energetic
than a normal supernova. Above $\sim 260\, M_{\odot}$ there is direct
collapse into a BH because nuclear burning is unable to halt
the collapse and generate an explosion. The remnant mass exceeds half
of the initial stellar mass, thus constituting an IMBH. Objects that
are initially more massive than $\sim 10^5\, M_{\odot}$ cannot have
stable hydrogen burning to begin with, due to a post-Newtonian
instability. As a result, such objects quickly collapse into a BH
(Baumgarte \& Shapiro 1999; Shibata \& Shapiro 2002).

It is thus possible, and maybe even likely, that a population
of IMBHs was produced from Population~III stars. The size of this
population was recently estimated by Madau \& Rees (2001) and
Schneider et al.~(2002). Many details of these calculations are
uncertain, but both papers find that a population could easily have
been produced with a global mass density similar to that of the
supermassive BHs in the Universe, and possibly more. The
IMBHs would presumably have formed at redshifts $z \approx 10$--20 in
peaks of the mass distribution.\looseness=-2

\subsection{Formation in Dense Star Clusters}
\label{ss:dense}

Star clusters have long been suspected as possible sites for the
formation of IMBHs. The self-gravity of a cluster gives it a negative
heat capacity that makes it vulnerable to the so-called ``gravothermal
catastrophe'': the core collapses on a timescale proportional to the
two-body relaxation time (Binney \& Tremaine 1987). The resulting high
central density may lead to BH formation in various ways. The crucial
issue is whether realistic initial conditions ever lead to densities
that are high enough for this to occur, or whether core-collapse is
halted and reversed at lower densities. This question has been
addressed theoretically using semi-analytic arguments, Fokker-Planck
calculations, and direct $N$-body codes.

Lee (1987) and Quinlan \& Shapiro (1990) studied the importance of
stellar mergers during core collapse. These can give rise to the
runaway growth of a supermassive star, which at the end of its
lifetime collapses to a BH (\S~\ref{ss:popIII}). These studies
found that runaway merging occurs naturally in very dense clusters
($\rho > 10^6\, M_{\odot} \> {\rm pc}^{-3}$) of many stars ($N >
10^7$). These initial conditions correspond to velocity dispersions of
hundreds of \kms, and may be relevant for (early) galactic
nuclei. This provides a scenario for the formation of an IMBH, which
through accretion might subsequently grow to become a supermassive
BH. By contrast, Quinlan \& Shapiro (1987, 1989) and Lee (1993)
studied the fate of a cluster of compact objects (neutron stars and
stellar-mass BHs) instead of normal stars. In this situation they
found, also for initial conditions appropriate for galactic nuclei,
that the core collapses all the way to a relativistic state. When the
redshift reaches values $z > 0.5$ (velocities in excess of $10^5$
\kms), a relativistic instability sets in that results in catastrophic
collapse to a BH (Shapiro \& Teukolsky 1985). However, this relativistic path
to a BH may not be the most natural evolutionary scenario.
Starting from a cluster of normal stars, runaway merging would likely
produce a single IMBH before a cluster of compact objects could form
(Quinlan \& Shapiro 1990).

The aforementioned studies agreed that formation of an IMBH would not
occur in star clusters with fewer than $10^6$--$10^7$ stars, such as
globular clusters. In such clusters core collapse is halted by binary
heating (Hut et al. 1992) before the densities become high enough for
runaway stellar merging. Three-body interactions between ``hard''
binaries and single stars add energy to the cluster (at the expenses
of the binaries, which become harder; Heggie 1975). The binaries form
primarily through tidal capture. This process is much more efficient
at the low velocity dispersions characteristic of globular clusters
than at the higher velocity dispersions of galactic
nuclei.

Although it has long been thought that core collapse in globular
clusters is generically halted by binary heating, it was realized
recently that this is not always true. Stars of different masses are
not always able to reach energy equipartition (Spitzer 1969), and in
fact, the Salpeter IMF is unstable in this sense (Vishniac 1978). This
causes the heaviest stars to undergo core collapse more or less
independently of the other cluster stars, on a timescale that is much
less than the core collapse time for the cluster as a whole. Portegies
Zwart \& McMillan (2002) used $N$-body simulations to show that a
runaway merger among these massive stars leads to the formation of
an IMBH, provided that the core collapse proceeds faster than their
main-sequence lifetime. This implies an initial half-mass relaxation
time $< 25$ Myr. For a globular cluster that evolves in the Galactic
tidal field the corresponding present-day half-mass relaxation time
would have to be $< 10^8$ yr. It has been proposed that this scenario
might be important for young compact star clusters, such as those
often observed in star-forming galaxies (Ebisuzaki et al. 2001). Many
of the Milky Way's globular clusters have half-mass relaxation times
in the range $10^8$--$10^9$ yr, and some have half-mass relaxation
times below $10^8$ yr (Harris 1996). So this scenario may well be
relevant for Milky Way globular clusters as well, in particular
because the physical conditions during their formation are only poorly
understood.

A more unlikely route for the formation of IMBHs in globular clusters
is through the repeated merging of compact objects, such as
stellar-mass BHs (Lee 1995; Taniguchi et al. 2000; Mouri \&
Taniguchi 2002). Such objects get caught in binaries through dynamical
effects. After hardening by interactions with single stars, they
eventually merge after losing energy by gravitational radiation.
However, the interactions that produce hardening also provide recoils
that tend to eject the binaries from the cluster (Kulkarni, Hut, \& McMillan 
1993; Sigurdsson \& Hernquist 1993; Portegies Zwart \&
McMillan 2000). This limits the scope for considerable growth through
repeated merging, although in some situations four-body interactions
may boost the probability (Miller \& Hamilton 2002a). One way to avoid
the recoil problem is to assume that there is a single BH somewhere in
the cluster that starts out at $\sim 50\, M_{\odot}$.  After sinking to
the cluster center through dynamical friction, the BH could slowly
grow in mass through merging with stellar-mass BHs. The mass of the BH
would be large enough to prevent ejection through recoil (Miller \&
Hamilton 2002b).

\subsection{Relation to Supermassive Black Hole Formation}
\label{ss:relatedsupBH}

The formation of supermassive BHs in the centers of galaxies is
poorly understood, but there are many plausible scenarios
(Begelman \& Rees 1978; Rees 1984). Many scenarios are extensions of
those discussed in the preceding sections, and involve IMBHs at some
time in their evolution. It is therefore possible that supermassive
BHs and IMBHs in the Universe are intimately linked. Also, not all BHs
in galaxy centers may have had the opportunity to become
supermassive, so some galaxies may have a central BH of
intermediate mass.

Scenarios that evolve IMBHs into supermassive BHs usually invoke
merging and/or accretion. Schneider et al.~(2002) and Volonteri,
Haardt, \& Madau (2003) considered the case of IMBHs formed from
Population~III stars (\S~\ref{ss:popIII}). They envisaged that
while galaxies are assembled hierarchically from smaller units, the
IMBHs in these units sink to the center through dynamical
friction. There they merge to form supermassive BHs. Haiman and Loeb
(2001) found that this is a plausible scenario for building some $10^9\,
M_{\odot}$ BHs at very early times, as required observationally by the
detection of bright quasars at redshifts as large as 6 (Fan et
al. 2001). However, Islam, Taylor, \& Silk (2003) argued that this
scenario may not be able to account for all the mass observed today in
supermassive BHs. Also, Hughes \& Blandford (2003) showed that
supermassive BHs that grow through mergers generally have little
spin, which makes it unlikely that such BHs could power radio jets.
As an alternative to merging, a single intermediate-mass ``seed'' BH
might have grown supermassive through accretion. The growth may
happen quickly through collapse of a surrounding protogalaxy onto the
BH (Adams, Graff, \& Richstone 2001) or it may happen slowly by
accretion of material shed by surrounding stars (Murphy, Cohn, \&
Durisen 1991). Feedback from the energy release near the center may
limit both the growth of the BH (Haehnelt, Natarajan, \& Rees 1998) and
the growth of the galaxy (Silk \& Rees 1998). Feedback from star
formation may also limit the BH growth (Burkert \& Silk 2001), while
fresh gas supply provided during the merging of galactic subunits
(Haehnelt \& Kauffmann 2000; Kauffmann \& Haehnelt 2000) may provide
increased growth. These scenarios can reproduce observed correlations
such as those between BH mass and bulge mass or bulge velocity
dispersion (\S~\ref{s:intro}).

Not all scenarios for supermassive BH formation proceed through an
IMBH stage. If a collapsing gas cloud can loose its angular momentum
and avoid fragmentation into stars, it may collapse to a BH
directly. Haehnelt \& Rees (1993) sketched a route by which this may
have occurred. Bromm \& Loeb (2003) investigated this possibility
quantitatively by studying the collapse of metal-free primordial
clouds of $10^8\, M_{\odot}$ using hydrodynamical simulations. To avoid
fragmentation into stars, they assumed that the presence of H$_2$
(which would otherwise be responsible for cooling) is suppressed by an
intergalactic UV background. With this assumption, condensations of
$\sim 5 \times 10^6\, M_{\odot}$ form that can collapse to a BH through
the post-Newtonian instability (\S~\ref{ss:popIII}). It is
unclear whether feedback from a growing BH may limit the attainable
mass, so it is possible that the result would be an IMBH.\looseness=-2

\subsection{Primordial Formation}
\label{ss:primordial}

BHs might have formed primordially in the early Universe. The mass of
such BHs is generally of order the horizon mass at its formation time,
$M \approx 10^5 (t/{\rm sec})\, M_{\odot}$ (Barrow \& Carr 1996),
although smaller values are not impossible (Hawke \& Stewart 2002).
At the Planck time ($\sim 10^{-43} {\rm sec}$) the horizon mass is the
Planck mass ($\sim 10^{-38}\, M_{\odot}$) and at $1 \> {\rm sec}$ it
is $10^{5}\, M_{\odot}$. Primorial BHs less massive than $\sim
10^{-18} \, M_{\odot}$ would by now have evaporated through the
process of Hawking radiation. Primordial BHs around this mass would
currently be evaporating, and the observed $\gamma$-ray background
places useful limits on their existence (MacGibbon \& Carr
1991). However, Hawking radiation becomes progressively less relevant
for more massive BHs, and is negligible for the mass range of interest
in the present context. So it places no useful observational limits on
the existence of primordial BHs in the intermediate-mass regime, and
our thinking must be guided by theoretical considerations.

One possible mechanism for primordial BH formation is through collapse
of density fluctuations (Carr 1975; Carr \& Lidsey 1993). However, in
standard cold dark matter (CDM) cosmologies the early Universe is
characterized by a very high degree of homogeneity and isotropy. The
associated Gaussian density fluctuations are much too small to
collapse to a BH (Begelman \& Rees 1998). Another mechanism for the
formation of primordial BHs does not require density fluctuations, but
invokes collisions of bubbles of broken symmetry during phase
transitions in the early Universe (Hawking, Moss \& Stewart 1982;
Rubin, Khlopov, \& Sakharov 2000). For example, the quantum chromodynamic 
cosmic phase transition at $t = 10^{-5} {\sec}$ might have produced
BHs of order $\sim 1\, M_{\odot}$ (Jedamzik 1997). Primordial BHs could
also have formed spontaneously through the collapse of cosmic strings
(MacGibbon, Brandenburger \& Wichowski 1998) or through inflationary
reheating (Garcia-Bellido \& Linde 1998). Even if primordial BH
formation is possible in these scenarios, it is by no means
guaranteed. Also, these scenarios do not naturally lead to BHs in the
intermediate-mass range. Afshordi, McDonald \& Spergel (2003) recently
addressed some cosmological implications of a potential large
population of primoridial IMBHs. However, more conventional
cosmological thinking suggests that such a population is not
particularly likely.

\section{IMBHs: The Missing Baryonic Dark Matter?}
\label{s:darkmatter}

There is now considerable evidence that the matter density of the
Universe is $\Omega_{\rm m} \equiv \rho_{\rm m} / \rho_{\rm crit}
\approx 0.3$, with an additional $\Omega_{\rm \Lambda} \approx 0.7$ in
a cosmological constant or ``dark energy.'' Comparison of Big Bang
nucleosynthesis calculations with the observed abundances of light
elements yields the baryon density: $\Omega_{\rm b} =
0.041 \pm 0.004$ (this number scales as $H_0^{-2}$, and $H_0 = 70$
\kms\ Mpc$^{-1}$ was assumed; Burles, Nollett, \& Turner 2001). The
non-zero value of $\Omega_{\rm m} - \Omega_{\rm b}$ indicates the
presence of non-baryonic dark matter, with some form
of CDM being the most popular candidate.  However, this is probably
not the only missing matter in the Universe.  A detailed inventory of
the visible baryonic matter adds up to a best guess of
only $\Omega_{\rm v} = 0.021$ (Persic \& Salucci 1992; Fukugita, Hogan,
\& Peebles 1998). Although this number can be stretched with various
assumptions, it does appear that half of the baryons in the Universe
are in some dark form.

Carr (1994) provided a general review of the candidates for, and
constraints on, the baryonic dark matter. The
hypothesis that it could be a population of IMBHs in the halos of
galaxies (Lacey \& Ostriker 1985) is of particular interest in the
present context. Such a population is constrained observationally by
the dynamical effects it would have on its environment
(\S~\ref{ss:dynamical}) and by its gravitational lensing
properties (\S~\ref{ss:lensing}). Additional constraints exist if
the IMBHs are assumed to have formed from Population~III stars
(\S~\ref{ss:popIIIremnants}).

\subsection{Dynamical Constraints on IMBHs in Dark Halos}
\label{ss:dynamical}

The gravitational interactions that one would expect IMBHs to have
with other objects provide important constraints on their possible
contribution to the baryonic dark matter (Carr \&
Sakellariadou 1999). For example, IMBHs in dark halos would heat
(increase the stellar velocity dispersion) of galaxy disks, the more
so for larger BH masses. The observed velocity dispersions of stellar
disks therefore limit the masses of BHs in galactic halos. If the
Milky Way dark halo were composed entirely of BHs, then their mass
would have to be less than $\sim 3 \times 10^6\, M_{\odot}$ (Carr \&
Sakellariadou 1999). This limit becomes more stringent for small, dark
matter dominated galaxies. Rix \& Lake (1993) find an upper limit of
$\sim 6 \times 10^3\, M_{\odot}$ for the Local Group galaxy GR8,
although this is open to debate (Tremaine \& Ostriker 1999). Halo
IMBHs also tend to disrupt stellar systems in galaxies, in particular
globular clusters (Carr \& Sakellariadou 1999). Klessen \& Burkert
(1995) found that this excludes the possibility that dark halos are
made up entirely of BHs more massive than $\sim 5 \times 10^4\,
M_{\odot}$. If there are more massive BHs, they would have to make up
a smaller fraction of the halo: no more than 2.5\%--5\% for BHs of $\sim
10^6\, M_{\odot}$ (Murali, Arras, \& Wasserman 2000). However, these
constraints are quite uncertain because it is unknown what the
properties of globular cluster systems were when they
formed. It could even be that disruption of clusters by IMBHs may have
played an essential role in the shaping of the present-day number and
mass distribution of globular clusters (Ostriker, Binney, \& Saha
1989).

IMBHs in a galactic halo sink to the center through dynamical
friction. If they merge and accumulate there, then the observed masses
of supermassive BHs in galaxy centers constrain the mass and number
of halo IMBHs (Carr \& Sakellariadou 1999). Xu \& Ostriker (1994)
modeled this in detail, taking into account the timescale for merging
through emission of gravitational radiation and the possibility of
slingshot ejection of BHs in three-body interactions. They found that
unacceptable build-up of a central object occurs only for a halo made
up of BHs more massive than $\sim 3 \times 10^6\, M_{\odot}$. Consistent
with this, Islam et al.~(2003) found in a study of a cosmologically
motivated population of IMBHs (remnants of Population~III stars) that
the build-up of central objects remains within observationally
acceptable limits.

\subsection{Lensing Constraints on IMBHs in Dark Halos}
\label{ss:lensing}

Massive compact halo objects (MACHOs) can produce gravitational
microlensing amplification of the intensity of background stars
(Paczy\'nski 1986). Several teams have monitored stars in the Large
Magellanic Cloud (LMC) for a number of years to search for such
signatures. For the specific case of the LMC, the average
characteristic timescale for microlensing events is $\sim 130
(M/M_{\odot})^{1/2}$ days. Hence, MACHOs are progressively more
difficult to detect for increasing masses. Still, Alcock et al.~(2001)
calculated that they should have been able detect $\sim 1$ event of
multi-year duration toward the LMC if the Milky Way dark halo were
made entirely of $100\, M_{\odot}$ MACHOs. By contrast, no LMC events
were detected with durations in excess of $\sim 130$ days. This rules
out that the entire halo is made of MACHOs in the range $0.15$--$30\,
M_{\odot}$ (95\% confidence), although contribution of a fraction
below $\sim 25$\% is allowed. Alcock et al.~(2000) did detect many
shorter duration events, from which it was concluded that $\sim 20$\%
of the Milky Way halo may be composed of compact objects with masses
in the range $0.15$--$0.9\, M_{\odot}$.

The possible contributions of compact objects to the dark halos of
other (more distant) galaxies are constrained by various gravitational
lensing effects as well. However, these do not yet place limits in the
intermediate-mass regime that improve upon what is already known from
Big Bang nucleosynthesis (Carr 1994).

\subsection{Constraints on IMBHs Formed from Population~III Stars}
\label{ss:popIIIremnants} 

The cosmic density of IMBHs that have formed from Population~III stars
(\S~\ref{ss:popIII}) is limited by additional constraints (Carr
1994). Stars with initial masses below $260\, M_{\odot}$ expel much of
their metals at the end of their lifetime, enriching the
interstellar medium (ISM). The fact that the enrichment must have
been less than the lowest metallicities observed in Population~I stars
($Z \approx 10^{-3}$) limits the cosmic mass density $\Omega$ in such
Population~III stars to no more than $10^{-4}$. The cosmic density of
stars more massive than $260\, M_{\odot}$ (and their IMBH remnants) is
not constrained by metallicity considerations because they do not end
their life in a supernova explosion. However, they might shed helium
before their ultimate collapse. This places some constraints on their
potential contribution to $\Omega$, but these are no more stringent
than what is already known from Big Bang nucleosynthesis. Objects more
massive than $10^5\, M_{\odot}$ shed neither metals nor helium because
they collapse to a BH without reaching stable hydrogen burning.

All Population~III stars below $10^5\, M_{\odot}$ shine brightly during
their main-sequence phase. Their numbers are therefore constrained by
observations of the extragalactic background light, particularly in
the infrared. The constraints depend on the formation redshift and on
whether or not allowance is made for possible reprocessing by
dust. Depending on the exact assumptions, a cosmic density of
Population~III stars sufficient to explain all the baryonic dark
matter may just barely be consistent with the available extragalactic
background light data (Carr 1994; Schneider et al. 2002). There are
also limits from the accretion that one would expect onto a
cosmologically important population of BHs, but these do not place
strong constraints in the intermediate-mass range (Ipser \& Price 1977;
Carr 1994).

\section{Searches for Individual IMBHs}
\label{s:obsconstraints}

IMBHs may contribute as much as $\Omega \approx 0.02$ to the cosmic
baryon budget (\S~\ref{s:darkmatter}). However, even if they
existed in far smaller numbers, they would be of great importance for
astrophysics. For comparison, the cosmic mass density of supermassive
BHs in galaxy centers is only $10^{-5.7}$ (Yu \& Tremaine
2002). Cosmologically motivated scenarios of Population~III evolution
easily predict densities that rival or exceed this (Madau \& Rees
2001; Schneider et al. 2002). Hence, it is important to search for
evidence of individual IMBHs. Such IMBHs may exist in the main
luminous bodies of galaxies, where they could reveal themselves
through their microlensing properties (\S~\ref{ss:bulgelensing}),
or through accretion-powered X-ray emission
(\S~\ref{ss:ULXs}). Alternatively, they could exist in the
centers of galaxies (\S~\ref{ss:galcenters}) or globular star
clusters (\S~\ref{ss:globclusters}). In the near future it may be
possible to search for gravitational-wave signatures of IMBHs
(\S~\ref{ss:gravwaves}).

\subsection{Bulge Microlensing}   
\label{ss:bulgelensing}

Compact objects can be detected through microlensing. The Einstein
ring crossing time scales as $M^{1/2}$ and long-duration events are
therefore of particular interest. No long-duration events were
detected toward the LMC (\S~\ref{ss:lensing}) but the situation
is different for the Galactic bulge: $\sim 10$\% of the few hundred
detected events have timescales exceeding $\sim 140$ days (Bennett et
al. 2002).

The lensing timescale depends not only on the lens mass, but also on
the unknown transverse velocity of the lens and the ratio of lens and
source distances. The latter quantities can be constrained
statistically from the fact that they are drawn from the known
phase-space distribution function of the Galaxy. When many events are
modeled as a statistical ensemble this yields an estimate of the mass
distribution of the lenses (Han \& Gould 1996). Additional information
is needed to constrain the masses of individual lenses. This is often
possible for long-duration events from the ``microlensing parallax''
effect, which produces a signature in the light curve due to the fact
that the Earth moves around the Sun as the event progresses. Modeling
yields the transverse velocity of the lens as projected to the solar
position. Bennett et al.~(2002) identified six events with
sufficiently accurate parallax data to yield an estimate of the lens mass. The
largest masses are $6^{+10}_{-3}\, M_{\odot}$ (MACHO-96-BLG-5) and
$6^{+7}_{-3}\, M_{\odot}$ (MACHO-98-BLG-6). The observational limits on
the lens brightness make these events excellent candidates for
stellar-mass BHs, the first tentative detections outside of XRBs. The
long-duration event MACHO-99-BLG-22/OGLE-1999-BUL-32 is even more
interesting (Agol et al. 2002; Mao et al. 2002). The lens-mass
likelihood function is bimodal, with maximum likelihood at mass
$130^{+42}_{-114}\, M_{\odot}$ and with a secondary peak of lower
likelihood at $4.0^{+1.5}_{-1.8}\, M_{\odot}$ (Bennett et al. 2003). So
this lens could be an IMBH.

The bulge microlensing events suggest that stellar-mass BHs may
contribute more than 1\% of the Milky Way mass (Bennett et al. 2002,
2003), more than is traditionally believed (Brown \& Bethe 1994; Fryer
1999). An important caveat in the analysis is that the dynamics of the
lenses is assumed to follow that of the known stars. This would be
violated if BHs are born with large kick velocities, as are neutron
stars. There is conflicting observational evidence on this issue
(Nelemans, Tauris, \& van den Heuvel 1999; Mirabel et al. 2002).

\subsection{Ultra-Luminous X-ray Sources}
\label{ss:ULXs}

The Eddington luminosity for an accreting compact object of mass $M$
is $1.3 \times 10^{38} \> (M/M_{\odot})$ \lum. This is $\sim 2 \times
10^{38}$ \lum\ for a neutron star and ($0.4$--$2) \times 10^{39}$ \lum\
for a stellar-mass BH. Surprisingly, X-ray observations with {\it Einstein}\
(Fabbiano 1989), {\it ROSAT}\ (Colbert \& Mushotzky 1999; Roberts \& Warwick
2000; Colbert \& Ptak 2002) and {\it Chandra}\ have shown that more luminous
sources appear to exist in $\sim 30$\% of nearby galaxies. If the
emission of these sources is assumed to be isotropic, then the
observed fluxes indicate X-ray luminosities $2 \times 10^{39} \leq L_{\rm X}
\leq 10^{41}$ \lum. These sources are generally referred to as
ultra-luminous X-ray sources (ULXs); their isotropic luminosities are
less that those of bright Seyfert galaxies ($10^{42}$--$10^{44}$ \lum), so
they are also sometimes referred to as ``intermediate-luminosity X-ray
objects.''

ULXs do not generally reside in the centers of galaxies, so they are
unrelated to low-level AGN activity. They are generally unresolved at
the high spatial resolution ($\sim$0\farcs5) of {\it Chandra}. Combined with
the fact that many show variability (Fabbiano et al. 2003), this rules
out the hypothesis that ULXs are closely spaced aggregates of
lower-luminosity sources. A detailed study of the ``Antennae''
galaxies (Zezas et al. 2002; Zezas \& Fabbiano 2002) shows that the
large majority do not have radio counterparts. Combined with the
observed variability, this rules out that they are young
supernovae. Hence, ULXs are believed to be powered by accretion onto a
compact object. Bondi accretion from a dense ISM is insufficient to
explain the observed luminosities (King et al. 2001), so the accretion
is believed to be from a companion star in a binary system. This
interpretation is supported by the variability seen in ULXs (in one
case there is even evidence for periodicity; Liu et al. 2002) and the
fact that some show transitions between hard and soft states (Kubota
et al. 2001). These characteristics are commonly seen in Galactic
XRBs.

If ULXs are emitting isotropically at the Eddington luminosity, then
the accreting objects must be IMBHs with masses in the range
$15$--$1000\, M_{\odot}$. Sub-Eddington accretion or partial emission
outside the X-ray band would imply even higher masses. However, the
mass cannot be more than $\sim 10^6\, M_{\odot}$, or else the BH would
have sunk to the galaxy center through dynamical friction
(\S~\ref{ss:dynamical}; Kaaret et al. 2001). Either way, the IMBH
interpretation of ULXs has several problems (King et al. 2001; Zezas
\& Fabbiano 2002). There is no known path of double-star evolution
that produces a binary of the required characteristics (King et
al. 2001). One would need to assume that the IMBH was born
isolated, and subsequently acquired a binary companion through tidal
capture in a dense environment. This predicts a one-to-one
correspondence between ULXs and star clusters, which is not
observed. In the Antennae, ULXs are often observed close to,
but not coincident with, star clusters. This suggests a scenario in
which ULXs are XRBs that have been ejected out of clusters through recoil
(Portegies Zwart \& McMillan 2000). This precludes an IMBH because the
mass would be too large for the binary to be ejected (Miller \&
Hamilton 2002b).

An alternative to the IMBH interpretation is that ULXs are an unusual
class of XRBs. One possibility is that radiation is emitted
anisotropically, so that the luminosity is overestimated when assumed
isotropic. Mild beaming (King et al. 2001) and a relativistic jet
(K\"ording, Falcke, \& Markoff 2002; Kaaret et al. 2003) have both been
proposed.  It is also possible that ULXs are in fact emitting at
super-Eddington rates (Begelman 2002; Grimm, Gilfanov, \& Sunyaev 2002).
Observations show that ULXs are often associated with actively
star-forming regions or galaxies. The brightest known source resides
in the starburst galaxy M82 (Matsushita et al. 2000; Kaaret et al. 2001; 
Matsumoto et al. 2001), and the merging Antennae galaxy pair
has the most known sources in a single system (18 above $10^{39}$
\lum). The association with young stellar populations suggests that
ULXs might be related to high-mass XRBs (where ``high-mass'' refers to
the companion). Indeed, optical counterparts reported for ULXs suggest
a young star cluster in one case (Goad et al. 2002) and a single
O-star in another case (Liu, Bregman, \& Seitzer 2002). However, ULXs
have also been identified in elliptical galaxies (Colbert \& Ptak
2002) and globular clusters (Angelini, Loewenstein, \& Mushotzky 2001;
Wu et al. 2002), which suggests an association with low-mass XRBs. So
it may be that the ULX population encompasses different types of
objects, possibly related to Milky Way sources like SS433 and
microquasars (King 2002). Outbursts with luminosities similar to those
of ULXs have indeed been reported for some Milky Way sources
(Revnivtsev et al. 2001; Grimm et al. 2002).
 
X-ray luminosity functions provide additional information. In both the
Antennae and the interacting pair NGC 4485/4490 (Roberts et al. 2002)
the luminosity function has constant slope across the luminosity
boundary that separates normal XRBs from ULXs. This is not expected if
the two classes formed through different evolutionary paths, and hence
does not support the IMBH interpretation. However, if normal XRBs and
ULXs differ only in beaming fraction then one would also have expected
a break in the luminosity function (Zezas \& Fabbiano
2002).

The X-ray spectra of ULXs are important as well. Most tend to have
hard spectra that are well fit by a so-called multi-color disk
black-body model (Makishima et al. 2000).  Others are equally well fit
by a single power law (Foschini et al. 2002; Roberts et al. 2002). These 
results are consistent with an association with accreting binaries. The good 
fits of accretion disk models suggest that the bulk of the emission is not
relativistically beamed (Zezas et al. 2002). The inner-disk
temperature in the models is of the order of $kT = 1$--2 keV. This is
similar to values observed in Galactic microquasars, and is larger
than what would naturally be expected for an IMBH (Makishima et
al. 2000). On the other hand, Miller et al.~(2003) recently found
strong evidence for soft components in {\it XMM-Newton}\ spectra of the two
ULXs in NGC 1313. These soft components are well fit with inner-disk
temperatures of $\sim 150$ eV. Temperature scales with mass as $T
\propto M^{-1/4}$, so this was interpreted as spectroscopic evidence
that at least in these ULXs the accreting object is an IMBH of $\sim
10^3\, M_{\odot}$.

\subsection{Galaxy Centers}
\label{ss:galcenters}

Dynamical studies of galaxies indicate that they generally have
central supermassive BHs and that the BH mass scales with the
velocity dispersion of the host spheroid as $M_{\rm BH} \propto
\sigma^4$ (\S~\ref{s:intro}). This result is based on data for
galaxies with Hubble types earlier than Sbc, $\sigma > 70$ \kms, and
$M_{\rm BH} > 2 \times 10^6\, M_{\odot}$. It is unknown whether the same
$M_{\rm BH}$--$\sigma$ relation holds for later-type and/or dwarf
galaxies. If so, then one would expect such galaxies to host IMBHs
(owing to their less massive spheroids and correspondingly smaller
velocity dispersions). However, no firm detections and mass
measurements exist for such galaxies. In fact, it is not guaranteed
that such galaxies have central BHs at all. This would provide a
natural explanation for the scarcity of AGNs among late-type
galaxies (Ho, Filippenko, \& Sargent 1997; Ulvestad \& Ho 2002). On the
other hand, we do know that at least some late-type galaxies host 
AGNs. The most famous example is NGC 4395, a dwarf galaxy of type Sm,
which has the nearest and lowest-luminosity Seyfert 1 nucleus yet
found (Filippenko \& Sargent 1989). The conventional explanation of
Seyfert activity suggests that at least this galaxy must have a
central BH. Filippenko \& Ho (2003) argue that the BH mass lies in the range 
$10^4-10^5\, M_{\odot}$, which puts it firmly in the intermediate-mass regime.

Dynamical measurements of BH masses in late-type and dwarf galaxies
are complicated by the fact that such galaxies generally host a
nuclear star cluster of mass $10^6$--$10^7\, M_{\odot}$ (B\"oker et
al. 2002). The cluster is often barely resolved at {\it Hubble Space
Telescope (HST)}\ resolution so that its gravitational influence
resembles that of a point mass. This masks the dynamical effect of any
BH, unless the BH is at least as massive as the cluster. This is not
expected in view of the $M_{\rm BH}$--$\sigma$ relation, and is indeed
generally ruled out by detailed modeling. B\"oker, van der Marel \&
Vacca (1999) inferred a BH mass upper limit of $5 \times 10^5\,
M_{\odot}$ for the nearby Scd spiral IC~342. Geha, Guhathakurta, \& van
der Marel (2002) inferred upper limits in the range $10^6$--$10^7\,
M_{\odot}$ for six dwarf elliptical galaxies in Virgo. The only way to
obtain more stringent limits is to study galaxies in the Local Group,
for which it is possible to obtain spectroscopic observations that
resolve the central star cluster itself. This was done for M33
($\sigma = 24$ \kms), with no resulting BH detection. Two independent
groups analyzed the same {\it HST}\ spectra, and obtained upper limits of
$1500\, M_{\odot}$ (Gebhardt et al. 2001) and $3000\, M_{\odot}$ (Merritt,
Ferrarese, \& Joseph 2001). This is a factor $\sim 10$ below the value
predicted by extrapolation of the $M_{\rm BH}$--$\sigma$ relation.

\subsection{Globular Clusters}
\label{ss:globclusters}

The existence of theoretical scenarios for IMBH formation in dense
star clusters (\S~\ref{ss:dense}) makes it natural to search for
IMBHs in globular clusters. This search splits into two questions:
does the mass-to-light ratio ($M/L$) increase toward the center in
globular clusters? (\S~\ref{sss:globML}); and can this be
explained as a result of normal mass segregation, or must an IMBH be
invoked? (\S~\ref{sss:segregate}).
  
\subsubsection{Centrally Peaked $M/L$ Profiles in Globular Clusters}
\label{sss:globML}

The radial $M/L$ profile of globular clusters is constrained by the
observed profile of the line-of-sight velocity dispersion $\sigma$
(through the equations of hydrostatic equilibrium). For distant
clusters one can use integrated light techniques similar to those used
for galaxy centers. Gebhardt, Rich, \& Ho (2002) performed such a study
for the globular cluster G1, the most massive cluster of M31. A
constant $M/L$ model cannot fit their {\it HST}\ data, and they inferred the
presence of $M_d = 2.0^{+1.4}_{-0.8} \times 10^4\, M_{\odot}$ of dark
material near the center. The corresponding ``sphere of influence'' $r_d
= G M_d / \sigma^2$ is only 0\farcs035, which is less than the {\it HST}\ 
FWHM. Nonetheless, it is plausible that $M_d = 2.0 \times 10^4\,
M_{\odot}$ can indeed be detected in G1: it is similar in distance and
physical properties to the central star cluster of M33, for which {\it HST}\ 
data yielded an upper limit as small as 1500--$3000\, M_{\odot}$
(\S~\ref{ss:galcenters}).

For Milky Way globular clusters, velocity determinations of individual
stars are better than integrated light techniques. The cluster M15 has
been observed from the ground by many groups (most recently by
Gebhardt et al. 2000) and has long been a focus of discussions on
IMBHs in globular clusters (as reviewed by van der Marel 2001). A
recent {\it HST}\ study (van der Marel et al. 2002) added important stars in
the central few arcsec of the cluster, yielding a combined sample of
$\sim 1800$ stars with known velocities. The inferred velocity
dispersion increases radially inward and cannot be fit with a
constant $M/L$ model. Gerssen et al.~(2002) modeled the data and
inferred the presence of $M_d = 3.2^{+2.2}_{-2.2} \times 10^3\,
M_{\odot}$ of dark material near the center.

There are $\sim 70$ pulsars known in globular clusters and some of
these have a negative period derivative ${\dot P}$. These can be used
to constrain the cluster mass distribution. Pulsars are expected to be
spinning down intrinsically (positive ${\dot P}$), so negative ${\dot
P}$ must be due to acceleration by the mean gravitational field.
This places a lower limit on the mass enclosed inside the
projected radius $R$ of the pulsar. In M15 there are two pulsars at $R
\approx 1''$ (Phinney 1993) whose negative ${\dot P}$ values are
consistent with the mass distribution implied by the stellar
kinematics (Gerssen et al. 2002).

D'Amico et al.~(2002) recently reported two pulsars with negative
${\dot P}$ at $6''$ and $7''$ from the center of the cluster NGC
6752. These suggest a large enclosed mass and a considerable central
increase in $M/L$. However, the inferred masses may be inconsistent
with the stellar kinematics of this cluster (Gebhardt 2002,
priv.~comm.). NGC 6752 is interesting also because it hosts a pulsar
at an unusually large distance from the cluster center. It has been
suggested that this pulsar may have been kicked there through
interaction with an IMBH in the cluster core (Colpi, Possenti, \&
Gualandris 2002; Colpi, Mapelli, \& Possenti 2003).

\subsubsection{IMBH versus Mass Segregation}
\label{sss:segregate}

A natural consequence of two-body relaxation in globular clusters is
mass segregation. In an attempt to reach equipartition of energy,
heavy stars and dark remnants sink to the center of the cluster, which
causes a central increase in $M/L$. One must model the time evolution
of the cluster in considerable detail to determine the theoretically
predicted $M/L$ increase. M15 is one of the few clusters for which
this has been done. The most recent and sophisticated Fokker-Planck
models constructed for M15 are those of Dull et al.~(1997). Gerssen et
al.~(2002) found that the $M/L$ profile published by Dull et
al.~(1997) did not contain enough dark remnants near the cluster
center to fit their {\it HST}\ data, which suggested the presence of an
IMBH. However, it was subsequently reported that the $M/L$ figure of
Dull et al. (1997) contained an error in the labeling of the axes (Dull et
al. 2003). A corrected data-model comparison shows that the
Fokker-Planck models can provide a statistically acceptable fit to the
{\it HST}\ data (Gerssen et al. 2003). Baumgardt et al.~(2003a) performed
direct $N$-body calculations and reached a similar conclusion.

Although models without an IMBH can fit the kinematical data for M15,
this does not necessarily mean that such models are the correct
interpretation. The dark remnants that segregate to the cluster center
evolve from stars with initial masses $M \geq 3\, M_{\odot}$. The
evolutionary end-products of such stars are only understood with
limited accuracy (Fryer 1999; Claver et al. 2001), and the same is
true for their IMF (especially at the low metallicities of globular
clusters). Depending on the assumptions that are made on these issues,
it is possible to create models that either do or do not fit the M15
data. Most of the neutron stars that form are expected to escape
because of kicks received at birth (Pfahl, Rappaport, \& Podsiadlowski
2002). The $M/L$ increase from mass segregation is therefore due
mostly to white dwarfs with masses $> 1\, M_{\odot}$. Such white dwarfs
have cooled for too long to be observable in globular clusters, which
makes this prediction hard to test. Another caveat is that M15 is
known to have considerable rotation near its center.  This is not
naturally explained by evolutionary models and may hold important new
clues to the structure of M15 (Gebhardt et al. 2000). Hence, an IMBH
of mass $\leq 2 \times 10^3\, M_{\odot}$ is certainly not ruled out in
M15 (Gerssen et al. 2003; Baumgardt et al. 2003a).

There are no X-rays observed from the center of M15. Ho, Terashima, \&
Okajima (2003) find $L_{\rm X}/L_{\rm Edd} \leq 4 \times 10^{-9}$. This does
not imply that there cannot be an IMBH. In globular clusters there is
only a limited gas supply available for accretion (Miller \& Hamilton
2002b) and an advection-dominated accretion flow (Narayan, Mahadevan
\& Quataert 1998) can naturally lead to very low values of $L_{\rm X}/L_{\rm
Edd}$. The galaxy M32, which has a well-established supermassive BH,
has an upper limit $L_{\rm X}/L_{\rm Edd} < 10^{-7}$ (van der Marel et
al. 1998).

Scaling of the Dull et al.~(1997, 2003) M15 models to the mass, size
and distance of G1 does not yield a sufficient concentration of dark
remnants to fit its {\it HST}\ data (Gebhardt 2002, priv.~comm.). Hence, G1
may well contain an IMBH in its center, as suggested by Gebhardt et
al.~(2002). If true, this need not necessarily be representative for
globular clusters in general. G1 is unusually massive, and it has been
suggested to be the nucleus of a disrupted dwarf galaxy (Meylan et
al. 2001). Either way, a simple scaling of the Dull et al.~models to
the case of G1 is likely to be an oversimplification. Baumgardt et
al.~(2003b) performed $N$-body calculations and argued that the G1
data can be explained without an IMBH. However, the proper scaling of
these calculations with $N \approx 7 \times 10^4$ particles to the
case of G1 (with $N \approx 10^7$ stars) is uncertain. The same
argument applies to the Baumgardt et al.~(2003a) models for M15; 
improved modeling of both clusters remains highly desirable.

It is intriguing that the BH mass detections/upper limits suggested
for G1 and M15 fall right on the $M_{\rm BH}$--$\sigma$ relation for
supermassive BHs. This leaves open the possibility that there may be
some previously unrecognized connection between the formation and
evolution of globular clusters, galaxies and central BHs. 

\subsection{Gravitational Waves}
\label{ss:gravwaves}

In the near future, gravitational wave detection experiments such as
LIGO and LISA will provide a new way to probe the possible existence
of IMBHs. Binary systems of compact objects and mergers of
supermassive BHs are already well known as possible sources of
gravitational radiation. Miller (2003) recently emphasized that a
population of IMBHs could also be observable, especially if they
reside in dense stars clusters. With optimistic assumptions, LIGO
could see the coalescence of a stellar-mass BH with an IMBH up to
several tens of times per year.

\section{Concluding Remarks}
\label{s:conc}

The main conclusion to emerge from this review is that the existence
of IMBHs in the Universe is not merely a remote possibility. IMBHs
have been predicted theoretically as a natural result of several
realistic scenarios. In addition, it has been shown that IMBHs might
plausibly explain a variety of recent observational findings. Much
progress has been made in the last few years, but certainly, even more
work remains to be done. None of the theoretical arguments for IMBH
formation are unique. Many alternative theoretical scenarios exist
that do not lead to IMBHs. Similarly, none of the observational
suggestions for IMBHs are clear cut. Alternative interpretations of
the data exist that invoke known classes of objects and many would
argue that such conservative interpretations are more
plausible. Either way, these issues can only be addressed and resolved
with additional research. IMBHs are therefore likely to grow as a
focus of research attention.

\end{document}